# Two states hydrogenlike model for High-Order Harmonic Generation and enhanced XUV Generation from a coherent superposition of bound states

S Batebi* and M Mohebbi
Department of Physics, Faculty of Science, University of Guilan, P. O. Box No: 41335-19141, Rasht, Iran.
E-mail: s_batebi@guilan.ac.ir

**Abstract.** We present an analytic, and fully quantum theory of high harmonic generation by low frequency laser fields for a two-state hydrogenlike atom model. The model is valid for an arbitrary atomic potential role of dipole matrix elements and can be generalized to describe laser fields of arbitrary polarization and includes the depletion of states. In the case of an initial coherent superposition of two states, the theory clearly explains why high conversion efficiency can be obtained. Our results are based on quantum interference effects in recombination via different states. We demonstrate that four quantum paths of electron contribute in the harmonic spectrum. The harmonic conversion efficiency is directly proportional to the initial-state amplitudes and depletion of the states (especially excited state), the remained population of the bound states at the time of recombination and the distinct dipole strengths. The main contribution to high conversion efficiency is the transition from the excited to the ground state.

## 1. Introduction

High-order harmonic generation HHG is a rapidly developing topic in the field of laser-atom interaction. Besides its fundamental interest, it represents an attractive technique for the production of ultrafast coherent radiation in the UV and soft-x-ray regions of the spectrum using a tabletop system. The HHG procedure can be well understood by the semiclassical three-step model [1] and the quantum model [2].

According to the three-step model, first the electron tunnels through the barrier formed by the Coulomb potential and the laser field, then it oscillates quasi freely driven by the laser field and acquires additional kinetic energy, and finally it can recombine with the parent ion and emit radiation. The harmonic spectrum is characterized by a rapid drop at low orders followed by a broad plateau where all the harmonics have the same strength and a sharp cut-off around harmonic energy $I_p + 3.17 U_p$, where $I_p$ is the atomic ionization potential and $U_p$ is the ponderomotive energy, i.e., the cycle-averaged kinetic energy of an electron gained in a mono chromatic laser field.

The main problem so far has been to generate sufficiently intense HHG pulses for practical applications. The most efficient methods depend on a single-atom signal, which is then phase matched in the target medium [3,4]. Most models of HHG are based on the Lewenstein model [2]. The Lewenstein model is often referred to as the strong-field approximation of HHG. In the above described models, the single atomic state was taken into account. We extend the Lewenstein model for the ground atomic state to two states case, analytically. To date, most of the theoretical work with the Lewenstein model, done on harmonic generation has focused on the response of atoms initially prepared in the ground state. This method does allow us to study some of the features that are unique for HHG from atoms or ions within a well-established framework. Examples of the importance of including the second state and underlying physics of the HHG, are preparing the initial state as a superposition of the ground and the first excited states or resonant interaction between bound atomic states and etc. We can simply apply the extracted result of our model and obtain the corresponding dipole momentum functions.

The driven two-level system is a particularly important model in physics, since it has proven to be very useful in describing many aspects of the interaction of matter with an electromagnetic field. By examining the populations of the ground and some excited state during the laser pulse, the conditions for achieving both high conversion efficiency and high cutoff frequency can be obtained. Before proceeding, we note that harmonic generation by preparing the initial state as a coherent superposition of two bound states was first proposed by Gauthey et all [5], Burnett and co-workers demonstrated that a harmonic spectrum with distinct plateaus could be obtained by such superposition states. The superposition state can be obtained by using one harmonic pulse with the frequency corresponding to the energy difference between the two bound states [6] before the fundamental laser pulse.

In particular, we want to show that if the initial state is prepared as a coherent superposition of the ground state and an excited state, it is possible to generate very high-order harmonics with high conversion efficiency using moderate laser intensities. By moderate intensities, we mean that the intensity is high enough to ionize the excited state, but the field is too weak to directly ionize the ground state. In this case, we found that the harmonic spectrum is composed of two distinct sets of harmonics.

This paper is organized as follows. Section 2 contains a general presentation of extended theory and a discussion of its quasi classical interpretation, supplemented by argument of the underlying physics. We present general expressions for harmonic strengths. We discuss the effects of depletion of the atomic system. In particular, for example, we study the case of 1s and 2s states. Section 3 presents the application of theory to Neon atom. Section 4 sums up our conclusions. Atomic units ($\hbar = e = m_e = a_0 = 1$) are used throughout, unless stated.

Since we aim to describe a rather general way of possible control over the harmonic emission, we perform our calculations for a simple hydrogenic atom, Ne. At this point it is worth pointing out that the results we present are not directly related to the structure of the atomic potential and therefore can be straight forwardly extended to any type of ion or atom.

## 2. Strong Field Approximation theory for two-state hydrogenlike atoms

Our point of departure is the treatment in [2], where the stationary-phase method is used to approximate the momentum integral (describing continuum dynamics). In this paper, we consider the interaction between an atom and a laser field in the single active electron approximation. The Lewenstein model, is a fully quantum mechanical approach which describes the HHG process. The basic equation of this theory is the time dependent Schrodinger equation (TDSE),

$$i\hbar \left| \dot{\psi}(t) \right\rangle = \hat{H} \left| \psi(t) \right\rangle,$$

$$\hat{H} = -\frac{1}{2m}\hat{\Pi}^2 + \hat{V}(r) - qE(t)\hat{x}$$

(1)

with the coulombic core potential operator $\hat{V}(r)$ and the dipole operator consisting of the laser electric field E (the x component) times the space operator $\hat{x}$. With the help of the TDSE one obtains the induced dipole moment:

$$x(t) = \left\langle \psi(\vec{r},t) \middle| \hat{x} \middle| \psi(\vec{r},t) \right\rangle,$$

(2)

which is directly linked to the high frequency spectra produced within the HHG process. In the following only linearly polarized light is considered, and the model is restricted to one electron ionization. Traditionally, this approach is confined to the case, where $I_p > \hbar\omega$ (typically $I_p \cong 5-20$ laser photons). One can go beyond this approximation by consideration resonant interaction between bound two atomic states. The ponderomotive potential $U_p$ should be in the range or larger than $I_p$ for ground state but still below the saturation energy $U_{sat}$, where all the gas atoms ionize during the interaction with the laser. For these parameters, tunneling theories, the ADK model [7], becomes valid. When the tunneling process emerges, the bounded electron is lifted up to a continuum state $|v\rangle$, denoted by its kinetic momentum.

Once it has appeared in the continuum, the interaction of the electron and the laser is determined by the last term of the Hamiltonian, $E(t)\hat{x}$. Assuming, that continuum to continuum (C-C) transitions do not contribute to the HHG process, the expectation value of the dipole moment with respect to transitions between to continuum states reads,

$$\langle v|\hat{x}|v'\rangle = -i\nabla_v \delta(v-v'). \tag{3}$$

In equation (1), the second term of the Hamiltonian represents the interaction of the electron and the atomic core potential $V(r)$. This term is negligible, because when the electron appears in the continuum it is immediately accelerated by the intense laser field. At the outer point of the electron trajectory, the kinetic energy of the electron is relatively low, but for these distances the core attraction vanishes also. Lastly, when the electron returns to the nucleus, it has gained such a high momentum, that again atomic potential forces can be neglected. This can formally be written as

$$\langle v'|\hat{V}(r)|\psi\rangle \approx 0. \tag{4}$$

The above considerations suggest that the following assumptions are valid:

(a) The contribution to the evolution of the system of all bound states except the ground $|0\rangle$ and the excited state $|1\rangle$ can be neglected.

(b) In the continuum, the electron can be treated as a free particle moving in the electric field with no effect caused by the atomic core potential $V(r)$.

Provided, that $U_p$ is large enough, (b) does not only hold for short range potentials, but also for long range potentials, as of hydrogen-like atoms. Assumption (b) implies that the electron has gained such a large kinetic energy when returning to the nucleus, that the atomic core attraction is negligible. This is not the case for lower harmonics of the order of $2q+1 \leq I_p/\hbar\omega$. Therefore, the Lewenstein model is only applicable to higher harmonics with photon energies $\hbar\omega_q \geq I_p$. In general, assumptions (a)-(b) are justified, if the so-called Keldysh parameter $\gamma = \sqrt{I_p/2U_q}$ is smaller than one, thus $I_p < 2U_q$.

In the following a formula, which approximately describes the induced dipole moment of the atom for the above discussed regime of parameters is derived by solving the time-dependent Schrodinger equation (1).

To solve this differential equation one has to make a conclusive ansatz. The considerations (a)-(b) motivate an approach, where the time-dependent wave function is split into two parts, the bound part represented by the ground state $a_1|0\rangle$ and some excited state denoted by $a_2|1\rangle$, and an integral over all unbound continuum states $b(v)|v\rangle$,

$$|\psi(t)\rangle = a_1(t)|0\rangle e^{-i\omega_1 t} + a_2(t)|1\rangle e^{-i\omega_2 t} + \int_0^\infty d^3v\, b(v,t)|v\rangle e^{-i\omega_v t}. \tag{5}$$

Thus, to solve equation (5) one has to find an expression for the amplitudes of the bound and continuum states $a_{1,2}$ and b. They are defined by a Schrodinger type equation,

$$\frac{\partial b(v,t)}{\partial t} = -iE(t)(a_1 d_1(v)e^{-i(\omega_1-\omega_v)t} + a_2 d_2(v)e^{-i(\omega_2-\omega_v)t})$$
$$- E(t)\frac{\partial b(v,t)}{\partial v} - i(\frac{1}{2}v^2 - \omega_v)b(v,t). \tag{6}$$

In the above equation $d_i(v) = \langle v|\hat{x}|i\rangle$ denotes the atomic dipole matrix element for bound-free transitions of electrons, which is later determined by the shape of the core potential. Thus, the whole information about the atom is reduced to the form of $d(v)$, and its complex conjugate $d^*(v)$. The differential

Schrodinger type equation (6) can be solved exactly, leading to a closed form of $b(v,t)$, using the laser $E(t)$ and the corresponding vector potential $A(t) = \int_{-\infty}^{t} E(t') dt'$:

$$b(v,t) = i\int_0^t dt' E(t') \{a_1(t') d_1(v + A(t) - A(t')) e^{-i(\omega_1 - \omega_v)t'}$$
$$+ a_2(t') d_2(v + A(t) - A(t')) e^{-i(\omega_2 - \omega_v)t'}\} \quad (7)$$
$$\times e^{-i\int_{t'}^t dt'' \{\frac{1}{2}(v+A(t)-A(t''))^2 - \omega_v\}}$$

Thus, the expectation of the position of the electron which is equal to the induced dipole moment, can be rewritten as

$$x(t) = \langle \psi(\vec{r},t) | \hat{x} | \psi(\vec{r},t) \rangle \approx \{a_2^* a_1 \langle 1|\hat{x}|0\rangle e^{-i(\omega_1-\omega_2)t} + a_1^* a_2 \langle 0|\hat{x}|1\rangle e^{-i(\omega_2-\omega_1)t}$$
$$+ \int d^3v\, b(v,t) a_1^*(t) d_1^*(v) e^{+i(\omega_1 - \omega_v)t} + \int d^3v\, b(v,t) a_2^*(t) d_2^*(v) e^{+i(\omega_2 - \omega_v)t}\} + C.C \quad (8a)$$

or

$$x(t) = \sum_{m \neq m'} \{a_m^* a_{m'} \langle m|\hat{x}|m'\rangle e^{-i(\omega_{m'} - \omega_m)t}\}$$
$$+ \sum_{m,m'} \{2\operatorname{Re}\{i a_m^*(t) e^{+i\omega_m t} \int d^3v\, d_m^*(v) \int_0^t dt' E(t') a_{m'}(t') \qquad (8b)$$
$$\times d_{m'}(v + A(t) - A(t')) e^{-i\omega_{m'} t'} e^{-i\int_{t'}^t dt'' \{\frac{1}{2}(v+A(t)-A(t''))^2\}}\}\}$$
$$\& \; m,m' = 1,2$$

Equations (8a) also indicate that dipole moments are directly related to the time-dependent amplitudes of the bound states. This is because harmonic generation originates from the coherent dipole transition between the continuum and the bound states.

In a next step a change of variables to the canonical momentum $p = v + A(t)$ is performed and the quasiclassical action is introduced.

The quasiclassical action

$$S(v,t,t') = \int_{t'}^t dt'' \{\frac{1}{2}(v + A(t) - A(t''))^2\} \quad (9)$$

describes the motion of a freely moving electron in the laser electric field. Hence an additional phase factor $e^{-iS(v,t,t')}$ enters the expression for the dipole moment. Note that equation (8) also takes into account some effects on the phase of the electronic wave packet due to the depth of the binding core potential (i.e. the ionization potential $I_P$). However, as it is just an additional constant, it neglects other perturbations of the electronic wave function due to the atomic potential $V(r)$. This means that the canonical momentum p is an observed quantity in between the time of ionization $t'$ and the time of recombination $t$. The expression for the induced dipole moment now reads as

$$x(t) \approx \sum_{m \neq m'} d_{mm'}^{b \leftrightarrow b} + \sum_{mm'} d_{mm'}^{b \leftrightarrow c}, \quad (10)$$

$$d_{mm'}^{b \leftrightarrow b} = a_m^* a_{m'} \langle m|\hat{x}|m'\rangle e^{-i(\omega_{m'} - \omega_m)t} \quad (10a)$$

$$d_{mm'}^{b \leftrightarrow c} = 2\operatorname{Re}\{i a_m^*(t) e^{+i\omega_m t} \int d^3P\, d_m^*(p - A(t)) \int_0^t dt' E(t')$$
$$\times a_{m'}(t') d_{m'}(p - A(t')) e^{-i\omega_{m'} t'} e^{-i\int_{t'}^t dt'' \{\frac{1}{2}(p-A(t''))^2\}}\} \& \; m,m' = 1,2 \quad (10b)$$

there is a rather intuitional physical interpretation of equation (10b) as continuous sum of probability amplitudes, which in principal corresponds to the three steps of HHG described in the Simple Man

Model. The first term in equation (10b), $E(t')d_{m'}(p - A(t'))$, represents the probability amplitude for the electron to perform a transition from the state $a_{m'}(t')e^{-i\omega_{m'}t'}|m'\rangle$ to the continuum at a time $t'$. During its flight, the electron is considered as a particle moving freely in the laser electric field and therefore the wave function describing the electron acquires an additional phase factor $e^{-iS(p,t,t')}$. The last term in equation (10b), the complex-conjugate of the dipole matrix element $d_m^*(p - A(t))$, can be interpreted as the probability amplitude for a recombination of the electron and the atomic core, to happen at the time $t$ together emission of a harmonic photon. The atom state is $a_m(t)e^{-i\omega_m t}|m\rangle$. An alternative, but slightly unintuitive interpretation of equation (10b) is a time-reversed process, where the electron appears in the continuum at a time $t$ in the future, propagates back to $t'$ and recombines with the nucleus. This interpretation allows for the invariance with respect to time included in the basic equation of this model, the TDSE (1).

In short speaking, the (10a) term represents a transition from the state $a_{m'}(t)e^{-i\omega_{m'}t}|m'\rangle$ to the state $a_m(t)e^{-i\omega_m t}|m\rangle$ at a time $t$. Physically, $d_{11}^{b\leftrightarrow c}$ and $d_{22}^{b\leftrightarrow c}$ are simply the dipole moments one would obtain starting in the ground and the excited states, respectively. On the other hand, $d_{12}^{b\leftrightarrow c}$ and $d_{21}^{b\leftrightarrow c}$ can be regarded as the interference between the two parts of the bound states (b), where the excited $|2\rangle$ (the ground state $|1\rangle$) is coupled to the continuum (c), inducing dipole moment between the continuum and ground states $|1\rangle$ (the excited state $|2\rangle$). This term is only relevant and significant if we start from a coherent superposition. An illustration of four quantum paths of electrons for recombination with parent ion together internal transition scratched schematically in figure 1.

The integral over the continuum states in equation (10b) can be approximately resolved via a stationary phase method, thus yielding a numerically computable expression for the second part of laser induced atomic dipole moment $d_{mm'}^{b\leftrightarrow c}$,

$$d_{mm'}^{b\leftrightarrow c} = 2\operatorname{Re}\{ia_m^*(t)\int_0^t dt'(\frac{\pi}{\varepsilon + i(t-t')/2})^{3/2} d_m^*(p_{st} - A(t))$$
$$E(t')a_{m'}(t')d_{m'}(p_{st} - A(t'))e^{-iS_{st}(t,t')}\} \qquad (11)$$

In the above equation, the linear polarization of the laser was taken in to account by writing $A(t)$ instead of the whole vector potential $\vec{A}$. Due to the integration method, the canonical momentum $p$ and the quasiclassical action $S(p,t,t')$ have been replaced by their stationary values

$$S_{st}(p_{st},t,t') = \omega_{m'}t' - \omega_m t + \int_{t'}^t dt''\{\frac{1}{2}(p_{st} - A(t''))^2\}, \qquad (12)$$

$$p_{st} = \frac{1}{t-t'}\int_{t'}^t dt'' A(t''). \qquad (13)$$

The additional prefactor in equation (11), $(\frac{\pi}{\varepsilon + i(t-t')/2})^{3/2}$, with an infinitesimal regularization constant $\varepsilon$, has arisen from the integration over all continuum states. It incorporates the spread of the electronic wave function due to quantum diffusion during the propagation in the continuum. Because of its proportionality to $(t-t')^{-3/2}$, it limits the electrons contributing to the HHG process to those only, which return after a few laser cycles.

The expression can be easily generalized to the case of laser fields $\vec{E}(t)$ of arbitrary polarization. If want to evaluate the component of the time-dependent dipole moment along the direction $\vec{n}$, where $\vec{n}$ is an unit vector, result is

$$\vec{r}(t) = \sum_{m \neq m'} \{a_m^* a_{m'} \langle m|\vec{r}|m'\rangle e^{-i(\omega_{m'}-\omega_m)t}\}$$
$$+ 2\text{Re} \sum_{m,m'} \{i a_m^*(t) \int_0^t dt' (\frac{\pi}{\varepsilon + i(t-t')/2})^{3/2} \vec{n}.\vec{d}_m^*(\vec{p}_{st} - \vec{A}(t)) \quad (14a)$$
$$\times \vec{E}(t').\vec{d}_{m'}(\vec{p}_{st} - \vec{A}(t')) a_{m'}(t') e^{-iS_{st}(\vec{p}_{st},t,t')}\} \quad \& m,m' = 1,2$$

$$S_{st}(p_{st},t,t') = \omega_{m'}t' - \omega_m t + \int_{t'}^t dt'' \{\frac{1}{2}(\vec{p}_{st} - \vec{A}(t''))^2\}, \quad (14b)$$

$$\vec{p}_{st} = \frac{1}{t-t'} \int_{t'}^t dt'' \vec{A}(t''). \quad (14c)$$

Here $a_m(t)$ are the time-dependent amplitudes of the ground and excited states, which is calculated by the ADK theory [8]. In the ADK theory, the instantaneous ionization rate for the $m$ th state in A.U. units is

$$W_m(t) = \omega_p |C_{n^*}|^2 (4\omega_p/\omega_t)^{2n^*-1} \exp(-4\omega_p/3\omega_t) \quad (15)$$

with

$$\omega_p = I_{p,m}, \omega_t = |\vec{E}(t)|/\sqrt{2I_p}, n^* = Z\sqrt{I_{ph}/I_{p,m}}$$
$$|C_{n^*}|^2 = 2^{2n^*}/(n^*\Gamma(n^*+1)\Gamma(n^*))$$

That $|\vec{E}(t)|$ is the amplitude of the electric field, Z the resulting net charge of the atom, $I_{p,m} = -\omega_m$ the ionization potential of the $m$ th state, $I_{ph}$ the ionization potential of hydrogen and $m_e$ electron mass. $\Gamma$ denotes the mathematical Gamma function. The ionization rate allows for an estimate of the probability of an atom to have remained in its ground state

$$R_m(t) = \exp[-\int_{-\infty}^t W_m(t')dt']. \quad (16)$$

Thus, the $a_m(t)$ term in equation (14) can be considered as $\sqrt{R_m(t)}$. Therefore, one should be cautious in employing tunnel rates in analyzing results for such laser parameters.

Finally, for the application of the above-presented theory, one should find the explicit form of the matrix element $\vec{d}_m$. For the case of hydrogenlike atoms and transitions from 1s and 2s states, the field-free dipole matrix elements can be approximated by [9]

$$\vec{d}_1(\vec{p}) = \langle \vec{p}|\vec{r}|0\rangle = i(\frac{2^{7/2}\alpha^{5/4}}{\pi})\frac{\vec{p}}{(\vec{p}^2+\alpha)^3} \text{ and } \vec{d}_2(\vec{p}) = \langle \vec{p}|\vec{r}|1\rangle = 4i(\frac{\alpha^{5/4}}{\pi})\frac{(\vec{p}^2-\alpha/2)\vec{p}}{(\vec{p}^2+\alpha/4)^3}, \text{ respectively, with}$$

$\alpha = 2I_p$. According to the selection rules, we get $\langle m|\vec{r}|m'\rangle = 0$.

That means that only the four terms of equation (14) will contribute to the harmonic generation. Neglecting the influence of the second bound state except the ground state and using the initial condition $a_m(t = -\infty) = \delta_{m,1}$, from equation (14) we obtain the well-known result in [2].

HHG power spectrum can be determined, which is proportional to the modulus squared of the Fourier transform of $\vec{r}(t)$. It is given by

$$\vec{P}(q\omega_0) = \left|\frac{1}{\sqrt{2\pi}} \int_0^t \vec{r}(t) e^{-iq\omega_0 t} dt\right|^2. \quad (17)$$

## 3. Results and discussion

To clearly understand the physical picture of the two-level systems in our scheme, we investigate the high-order harmonic generation process with arbitrary parameters; however, the purpose of this work is to show the general nature of the two-level systems with a generalized SFA model for single atom based on Lewenstein model, not to simulate any particular experiments.

Throughout this study, a Gaussian pulse and linearly polarized laser field is used. The electric field of the laser pulse is expressed by

$$E(t) = E_0 \cos(\omega_0 t) f(t) \qquad (18)$$

where $E_0$ is the peak amplitude of the electric field of the laser pulse, $\omega_0$ is the frequency of the 800 nm pulse. $f(t) = \exp(-4\ln(2)t^2/\tau_p^2)$ is the pulse envelope. The time evolves from 0 to $T = 2\tau_p$ that $\tau_p$ is the corresponding pulse duration $\tau_p = 10 fs$ (FWHM). In our simulation, the model atom is chosen to be Neon, and the ground and excited states binding energy are 15.76ev and 15.76/4 ev, respectively.

Let us consider an initial state prepared in such a way that there is maximal initial coherence between the two levels $|1\rangle$ and $|2\rangle$, by the action of a preparation pulse, applied prior to t=0 which is the turn-on time of the main pulse, so that (with equally weighted population)

$$|\psi(t=0)\rangle = \frac{|0\rangle + |1\rangle}{\sqrt{2}}. \qquad (19)$$

In this article, the laser intensity restricted to the validity limit of ADK regions. Based on the above argument, the results of equations (11-18) investigated in followed figures 2-4. We first present results for an intensity of $I_0 = 1.0e^{+13}$ W/cm$^2$.

The solid curves in figure 2a show the HHG spectra for a coherent superposition state as a function of the harmonic order for $d_{12}^{b\leftrightarrow c}$ (magenta solid line), $d_{21}^{b\leftrightarrow c}$ (blue dashed line) and $d_i = d_{12}^{b\leftrightarrow c} + d_{21}^{b\leftrightarrow c}$ (black dotted line) terms. Total term $d_{12}^{b\leftrightarrow c} + d_{21}^{b\leftrightarrow c}$ exhibits interference structure in the spectra. Physically, this spectral modulation should correspond to the interference between the harmonics in $d_{12}^{b\leftrightarrow c}$ and $d_{21}^{b\leftrightarrow c}$ terms of equation (10b), i.e. the crossed term of $\left|d_i^{b\leftrightarrow c}\right|^2 = \left|d_{12}^{b\leftrightarrow c} + d_{21}^{b\leftrightarrow c}\right|^2$. Consider the power spectra of the harmonics: $P_{12} = \left|d_{12}^{b\leftrightarrow c}\right|^2$, $P_{21} = \left|d_{21}^{b\leftrightarrow c}\right|^2$ and mixture of them $P_{12+21} = \left|d_{12}^{b\leftrightarrow c} + d_{21}^{b\leftrightarrow c}\right|^2$. As can be seen in the simplest case of two point emitters at the site 1 and 2,

$$\left|d_{12}^{b\leftrightarrow c} + d_{21}^{b\leftrightarrow c}\right|^2 = \left|d_{12}^{b\leftrightarrow c}\right|^2 + \left|d_{21}^{b\leftrightarrow c}\right|^2 + 2\text{Re}[d_{12}^{b\leftrightarrow c} \cdot d_{21}^{*b\leftrightarrow c}] \qquad (20)$$

where $d_i^{b\leftrightarrow c}$ is the harmonic amplitude at the site $i$, the effects of the interference between two terms (i.e., $2\text{Re}[d_{12} d_{21}^*]$) should appear in the difference between $P_{12+21}$ and $P_{21} + P_{12}$. The inset of figure 2a shows the focusing of $\left|d_i^{b\leftrightarrow c}(\omega)\right|^2$ around of the 7th and 17th harmonics. The harmonics around 17th are enhanced, and those around 7th are suppressed. It should be emphasized that $P_{12+21}(\omega_7)$ is completely suppressed compared to $P_{21}(\omega_7) + P_{12}(\omega_7)$. In fact, this spectral modulation should correspond to the interference of two trajectories together, from four trajectories (transition) between the ground and excited states, i.e. the whole of the various contributions to the dipole momentum for the coherent superposition as:

$$|d(t)|^2 = \left|d_{11}^{b\leftrightarrow c} + d_{22}^{b\leftrightarrow c} + d_{12}^{b\leftrightarrow c} + d_{21}^{b\leftrightarrow c}\right|^2 = \left|d_{11}^{b\leftrightarrow c}\right|^2 + \left|d_{22}^{b\leftrightarrow c}\right|^2 + \left|d_{12}^{b\leftrightarrow c}\right|^2 + \left|d_{21}^{b\leftrightarrow c}\right|^2 + \cdots \qquad (21a)$$

or

$$|d(t)|^2 = \left|d_{11}^{b\leftrightarrow c} + d_{22}^{b\leftrightarrow c}\right|^2 + \left|d_{12}^{b\leftrightarrow c} + d_{21}^{b\leftrightarrow c}\right|^2 + 2\operatorname{Re}[(d_{11}^{b\leftrightarrow c} + d_{22}^{b\leftrightarrow c}).(d_{12}^{b\leftrightarrow c} + d_{21}^{b\leftrightarrow c})^*]. \qquad (21b)$$

The first two terms on the right-hand side of equation (21b) are simply the square of dipole momentum (power spectra) one would obtain starting in the ground and excited states $P_{11+22} = \left|d_{11}^{b\leftrightarrow c} + d_{22}^{b\leftrightarrow c}\right|^2$ (as $\left|d_{1+2}\right|^2$ by the blue dashed line in figures 2,4(b)), and the crossed term $P_{12+21} = \left|d_{12}^{b\leftrightarrow c} + d_{21}^{b\leftrightarrow c}\right|^2$ (as $\left|d_i\right|^2$ by the red dotted line in figure 2,4(b)), respectively.

We see that the third term on the right-hand side of equation (21b), can be thought of as an interference between the first two terms of the total harmonic intensity $|d(\omega)|^2$ (figure 2-4(b)). We also see in inset of figure 4b, obviously that the third term in equation (21), can be thought of as an interference between the first two terms of the total harmonic intensity $|d(\omega)|^2$.

The first term in equation (21b), i.e. $P_{11+22}$ is negligible in comparison with the second term $P_{12+21}$ in plateau (only in figures 2,3(b)). Therefore, the term $d_i = d_{12}^{b\leftrightarrow c} + d_{21}^{b\leftrightarrow c}$ is, as we see, responsible for the plateau [as it can be seen in figure 2b.]. The HHG spectra of the superposition case in figure 2c (green solid line) shows only one plateau, which are about seven, ten orders of magnitude higher than that of the ground and excited states contributions, $d_1^{b\leftrightarrow c}$ and $d_2^{b\leftrightarrow c}$, blue dashed and red dash-dot lines, respectively (no generation of HOH by the ground and excited states contributions, $d_1^{b\leftrightarrow c}$ and $d_2^{b\leftrightarrow c}$, respectively). On the other hand, the plateau in the spectrum is due to the interference terms, $d_{12}^{b\leftrightarrow c}$ and $d_{21}^{b\leftrightarrow c}$ (magenta dotted line in figure 2c). The solid curves in figure 2d show the populations of the ground (blue dashed line), second excited (red solid line) states as a function of time (optical cycle). For this choice of intensity and wavelength the ionization of the ground state is negligible and there is only transference of population from the excited state to the continuum $|b| = \sqrt{1 - |a_1|^2 - |a_2|^2}$.

For the sake of comparison, we also display in figure 2e the harmonic spectra corresponding to an initial state of the ground state alone, $a_1(t=0) = 1$, $a_2(t=0) = 0$ (blue dashed line). The harmonic spectra of the superposition state (black solid line) clearly shows about seven - ten orders of magnitude higher than that of the ground state alone (blue dashed line) in figure 2e. Therefore, if the initial state is prepared as a coherent superposition of the ground state and an excited state, a large dipole transition will be induced between the continuum and bound states, i.e. the crossed trajectories, $d_{12}^{b\leftrightarrow c}$ and $d_{21}^{b\leftrightarrow c}$ due to high population of the continuum $b$ (equation (8)).

In figures 3,4, a comparison similar to the one case presented in figure 2, but for the higher laser intensities, I=(1-10)e$^{+14}$W/cm$^2$ are shown.

We now turn to the case of figure 3, where using the superposition state as the initial state significantly enhances the conversion efficiency of the high harmonics. As shown in figures, while the ground state population remains almost constant, the first excited state is ionized effectively in two optical cycles, leading to high populations of the continuum states.

In this case, the spectrum of the superposition is four orders of magnitude higher than that of the ground state case and about one order of magnitude higher than of the first excited state case. As clearly shown in figures 2,3, One can see that $|d| \approx |d_i|$ and $|d_i| > |d_2| > |d_1|$. All of these features can be easily understood from equation (8) and the following discussion.

Finally, we consider the strong field case. As shown in figure 4, the population of the first excited state decreases from 0.5 to 0.001 in about one optical cycles when the laser intensity reaches to $I_0=1.0e^{+15}W/cm^2$. One direct consequence of this dramatic ionization of the excited state is that, for $d_{22}^{b\leftrightarrow c}$ case in figure 4c (and figure 3c), the spectrum of these cases exhibits a monotonic decrease of the harmonic intensity (red dash-dot line). The spectrum of the superposition case exhibits a complex structure which has two plateaus (figure 4(b,c,e)).

We can see why this should be so by splitting up the various contributions to the dipole momentum for the coherent superposition. The first plateau increases by about one order of magnitude compared to that of the ground state case due to the interference contribution $|d_{12}|$ (magenta solid line in figure 4a and as $|d_i|^2$ magenta dotted line in figure 4c), while the other part of the spectrum agrees with that of the ground state case (blue dashed line in figure 4c). By comparing the first ground state spectrum alone (in figure 4(c,e)) with that of the superposition state case, we conclude that the intensity of the second plateau in the superposition case is the result of the contribution of the ground state (in agreement with our observations in figure 4c that indicated with $d_1$ red dash-dot line).

We can gain further insight into our interpretation of these expressions by comparing the harmonic spectra obtained directly from the dipole momentum contributions, i.e. $d_{mm'}$ with together. It is possible to generalize the approach of Ref. [2,10] (see also [11]) to the case when initially the ground and excited states have nonvanishing probability amplitudes $a(t)$ and $b(t)$.

To obtain the harmonic spectrum we have to calculate the Fourier transform of the various contributions in the dipole momentum of equation (11).

A good estimate of the integral over $t, t'(or \tau = t - t')$ can be obtained using the method of stationary phase [11].

To do that we replace the integral by a sum of contributions corresponding to stationary points of the Legendre transformed quasiclassical action

$$S_{st}(p_{st},t,t') - q\omega_0 t = \int_{t-\tau}^{t} dt'' \{\frac{1}{2}(p_{st} - A(t''))^2\} + \omega_{m'}(t-\tau) - \omega_m t - q\omega_0 t \qquad (22)$$

where q denotes the corresponding harmonic orders.

To compare the plateau levels, i.e. the ratio $P_{mm'}(l'\omega_0)/P_{nn'}(l\omega_0)$ ( $P_{mm'}(\omega) \propto |d_{mm'}^{b\leftrightarrow c}(\omega)|^2$ : the power spectrum), we take $l' = l$, where $l'$ and $l$ denote the corresponding harmonic orders. That means that the same electron trajectories provide the stationary points of the action (22). As a result we obtain

$$\frac{P_{mm'}}{P_{nn'}} \approx \left|\frac{a_m(t)}{a_n(t)} \frac{a_{m'}(t-\tau_s)}{a_{n'}(t-\tau_s)}\right|^2 \left|\frac{d_{m'}(p_{st} - A(t-\tau_s))d_m(p_{st} - A(t))}{d_{n'}(p_{st} - A(t-\tau_s))d_n(p_{st} - A(t))}\right|^2 \qquad (23)$$

where the field-free dipole moments are calculated at the stationary points, i.e., for the continuum state corresponding to the electron returning to the nucleus at time t with the appropriate velocity $P_{st} - A(t)$.

Thus, a ratio for the plateaus' conversion efficiencies directly related to the distinct dipole strengths, the amplitudes of the bound states.

To proceed further we limit our attention to the case when the kinetic energy of the returning electron $(p - A(t))^2/2$ is much larger than the ground and excited states binding energy. In particular we consider the harmonics at the end of the plateaus.

From the stationary point equations for the action (22), $\partial(S - q\omega_0 t)/\partial t = 0$ and $\partial(S - q\omega_0 t)/\partial \tau = 0$, we obtain

$$(p - A(t-\tau))^2 = -2I_{p,m'}$$
$$(p - A(t))^2 - (p - A(t-\tau))^2 + 2(\omega_{m'} - \omega_m) - 2q\omega_0 = 0 \Rightarrow (p - A(t))^2 = -2I_{p,m} + 2q\omega_0 \qquad (24)$$

So that

$$\frac{d_1(p(t)) = \langle p(t)|\hat{r}|0\rangle}{d_2(p(t)) = \langle p(t)|\hat{r}|1\rangle}$$
$$= \frac{2^{5/2}((p-A(t))^2 + I_{p,1}/2)^4}{((p-A(t))^2 + 2I_{p,1})^3((p-A(t))^2 - I_{p,1})} = \frac{2^{5/2}[-2I_{p,m} + 2q\omega_0 + I_{p,1}/2]^4}{[-2I_{p,m} + 2q\omega_0 + 2I_{p,1}]^3[-2I_{p,m} + 2q\omega_0 - I_{p,1}]} \qquad (25a)$$

$$\frac{d_1(p,t-\tau) = \langle p(t-\tau)|\hat{r}|0\rangle}{d_2(p,t-\tau) = \langle p(t-\tau)|\hat{r}|1\rangle}$$
$$= \frac{2^{5/2}((p-A(t-\tau))^2 + I_{p,1}/2)^4}{((p-A(t-\tau))^2 + 2I_{p,1})^3((p-A(t-\tau))^2 - I_{p,1})} = \frac{2^{5/2}[-2I_{p,m'} + I_{p,1}/2]^4}{[-2I_{p,m'} + 2I_{p,1}]^3[-2I_{p,m'} - I_{p,1}]} \qquad (25b)$$

where $I_{p,m} = -\omega_m$ is the ionization potential of the $m$ th state.

Approximately, for large $q$ values, the ratio between the matrix elements involved in the coherent superposition behaves then as

$$\frac{d_1(p(t))}{d_2(p(t))} \approx 2^{3/2} > 1, \quad \frac{d_1(p(t-\tau))}{d_2(p(t-\tau))} \approx 2^{-1/2} < 1. \qquad (26)$$

As a result, the probability of transition from the continuum back to the ground state $|1\rangle$ is higher than that to the excited state $|2\rangle$ while the transition from the ground state $|1\rangle$ to the continuum is lower than that the excited state $|2\rangle$ to the one i.e., equation (10b), above discussion about the continuous sum of probability amplitudes.

Therefore, a ratio for the plateaus conversion efficiencies (equation (23)) between the ground and excited states reduces to

$$\frac{P_{11}}{P_{22}} \approx 2\left|\frac{a_1^*(t)}{a_2^*(t)}\right|^2 \left|\frac{a_1(t-\tau_s)}{a_2(t-\tau_s)}\right|^2 \qquad (27)$$

Moreover, as shown in figure 4d, the laser intensity is high enough to ionize the excited state, while too weak to ionize the ground state. Since $|a_1| > |a_2|$ at the time of recombination, we have from equation (27) that $P_{11} > P_{22}$.

The same argument and proof as above lead to

$$\frac{P_{12}}{P_{21}} \approx \left|\frac{a_1(t)}{a_1(t-\tau_s)}\right|^2 \left|\frac{a_2(t-\tau_s)}{a_2(t)}\right|^2 \approx \left|\frac{a_2(t-\tau_s)}{a_2(t)}\right|^2 \qquad (28)$$

Since the amplitude of the ground state is approximately constant during the laser pulse $|a_1(t-\tau_s)| \approx |a_1(t)|$ and the population of the excited state decreases significantly within one optical cycle $|a_2(t-\tau_s)| > |a_2(t)|$ at the time of recombination, we have $P_{12} > P_{21}$ (figures 3,4(a)).

Similar considerations lead to the following expression for large $q$ values in plateau (figure 4(a,c)):

$$P_{11} > P_{12} > P_{22}. \qquad (29)$$

All of these results suggest that the relation (23 and 25) holds in practice, in excellent agreement with our numerical observations (figures 2,3,4).

From equation (29) and figure 4(a,c) one can deduce that the first and second plateau observed in the harmonic spectra simply corresponds to the $P_{12}$, $P_{11}$ contributions, respectively.

The presented results in figures 2,3, i.e., the low-intensity region, can be described in a similar way and become $P_{12} > P_{21}$. Thus the harmonics of the superposition case originate from the recombination in to the ground and excited states of the electron, $P_{12}$ and $P_{21}$ respectively (no generation of HOH by the ground and excited states contributions, $d_1^{b\leftrightarrow c}$ and $d_2^{b\leftrightarrow c}$, respectively) while the main contribution to the harmonic generation in this case of figure 4 is the transition from the continuum to the ground state i.e., $P_{12}$.

We are interested in producing harmonic photons with high conversion efficiency, which is directly proportional to the population of the continuum and the remaining population of the bound states, $b$ (i.e., $a_m(t')$) and $a_m^*(t)$ terms, respectively in equation (8). As a result, only those states that remain populated during the pulse will contribute to the harmonic generation (equations (23,28)) that gives $a_2^*(t)a_1(t')$ i.e., $P_{12}$. Hence, the conditions for high conversion efficiency of HHG, i.e. high populations for both the continuum and the ground state, are satisfied.

Therefore, in general there are three main factors which can affect the conversion efficiency of HHG, i.e., the initial-state amplitudes and depletion of the states (especially excited state)), the remained population of the bound states at the time of recombination and the distinct dipole strengths.

## 4. Conclusions

The conventional strong-field approximation based on Lewenstein model is not expected to be valid for initially coherent state preparing of atoms or ions due to the interaction between the two orbitals (levels) and the driving field. Analytically, the Lewenstein model is extended to two states with the depletion effect and resonance capability. The advantages of this approach are that it eliminates the costly calculation of numerical integration of the time dependent schrodinger equation for a hydrogenlike atom in a laser pulse, and accurately describes the physical picture of HHG from a two-level system. Our results are based on quantum interference effects in recombination via different states. Clearly, we have extracted that four quantum paths of electron contribute in the harmonic spectrum. In conclusion, we have demonstrated why the harmonic plateau enhanced with higher efficiencies than it is possible to obtain starting the process from the ground or excited state. Therefore, there are three main factors which can affect the conversion efficiency of HHG, i.e., the initial-state amplitudes and depletion of the states (especially excited state)), the remained population of the bound states at the time of recombination and the distinct dipole strengths.

The main contribution to high conversion efficiency is the transition from the excited to the ground state i.e., the advantage of using coherent superposition state is that it is possible to induce dipole transition between the continuum and the ground state, where the excited state is responsible for the ionization, thus, drastically increasing the conversion efficiency.

It is found that, the degree of the ionization of the states (the effect of the depletion) is important in the ratio of the transition contributions in the power spectrum. We demonstrate in particular that preparing the initial state in a coherent superposition of bound states leads to a harmonic spectrum with distinct plateaus with different conversion efficiencies that the transition from the excited to the ground state gives rise to the first plateau, while the transition from the ground to the ground state gives rise to the second one.

Our model reduces to the standard Lewenstein model for one state atom. Our approach allows for a simple analysis of harmonic efficiency as a function of a different laser and atomic parameters by superposition of electron paths. Finally, we should emphasize that the approach developed in the present paper can easily be generalized to multi-level atoms with resonance effects.

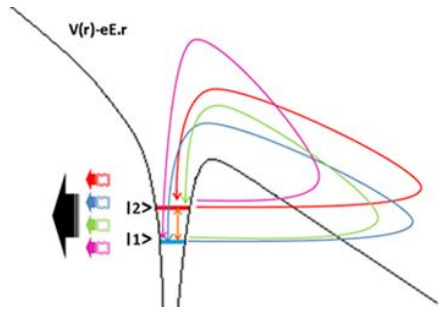

**Figure 1.** a simple picture of three-step processes;

Four quantum paths of electrons contribute the HHG for two levels atom.

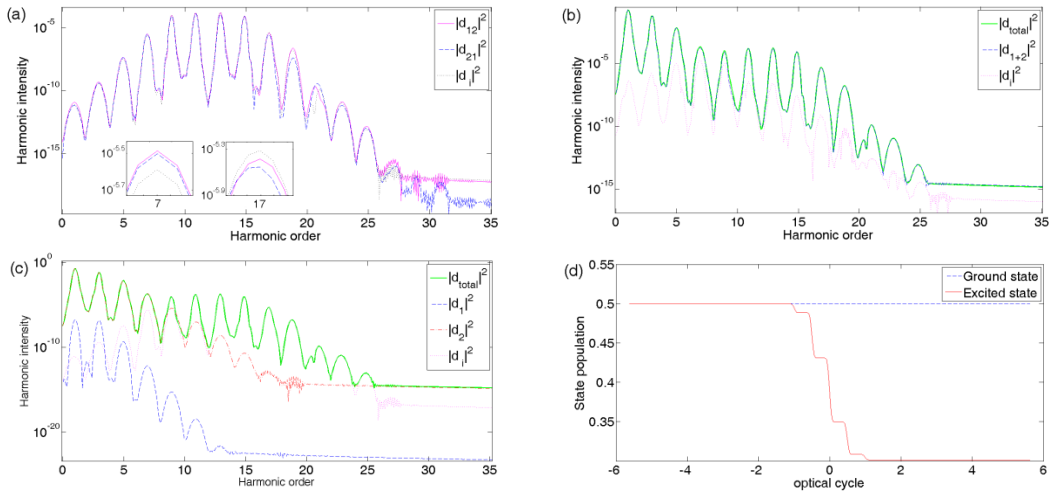

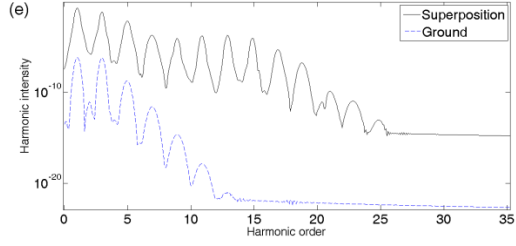

**Figure 2.** Harmonic generation spectra from the Neon. The pulse parameters is ( $\tau_p = 10fs$, $\lambda = 800nm$, $I_0 = 1.0e^{13}W/cm^2$ ). (a) interference terms $d_{12}^{b\leftrightarrow c}$ (magenta solid line), $d_{21}^{b\leftrightarrow c}$ (blue dashed line) and total interference terms, $d_{12}^{b\leftrightarrow c} + d_{21}^{b\leftrightarrow c}$ (black dotted line). Inset shows the enlarged illustration of the spectra, where the harmonics around 7th are enhanced, and those around 17th are suppressed. (b) $d$ (green solid line), $d_{1+2}^{b\leftrightarrow c}$ (blue dashed line), $d_i^{b\leftrightarrow c}$ (red dotted line). (c) $d$ (green solid line), $d_{11}^{b\leftrightarrow c}$ (blue dashed line), $d_{22}^{b\leftrightarrow c}$ (red dash-dot line), $(d_{12}^{b\leftrightarrow c} + d_{21}^{b\leftrightarrow c})$ (magenta dotted line). (d) Populations of the ground (blue dashed line) and second excited (red solid line) states as a function of time when the initial state is a coherent superposition of the ground and excited states with equally weighted populations. (e) Harmonic spectra corresponding to the ground state alone (blue dashed line), and coherent superposition state (black solid line).

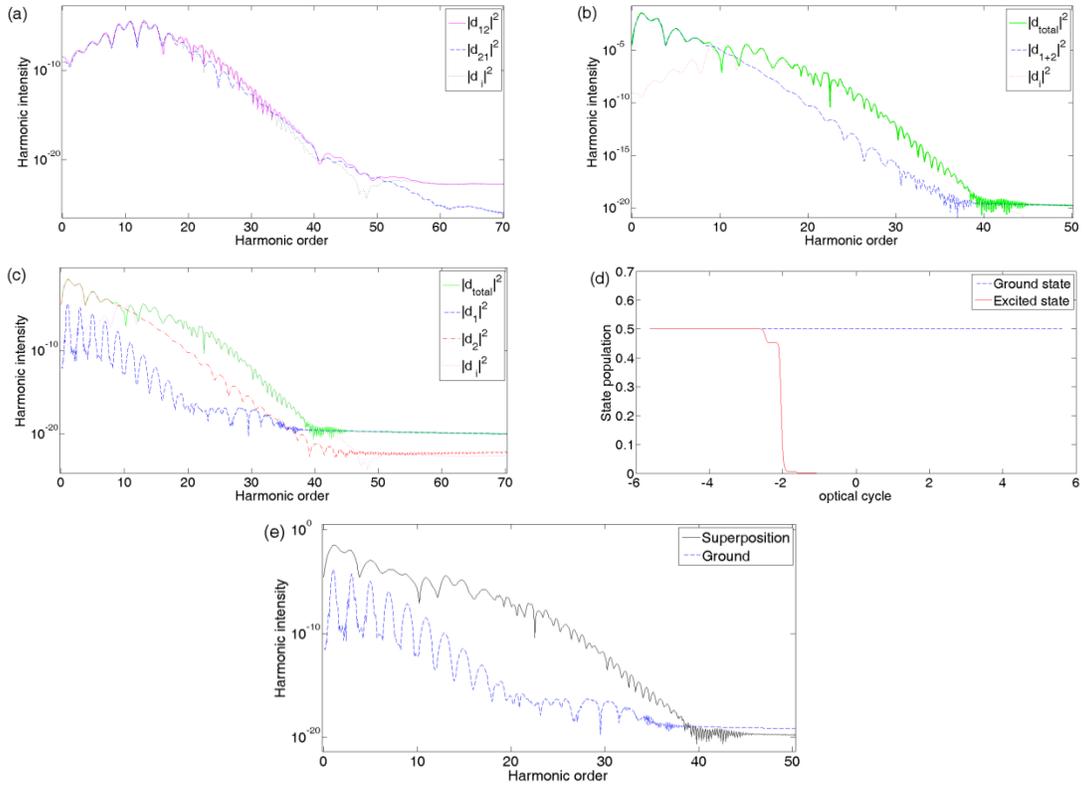

**Figure 3.** Conversion efficiency comparison between the various contributions to the dipole momentum, corresponding to the parameters of figure 2 by using the laser intensity at $I_0 = 1.0e^{14}W/cm^2$. (a) Harmonic intensity as a function of the harmonic order for interference terms $d_{12}^{b\leftrightarrow c}$ (magenta solid line), $d_{21}^{b\leftrightarrow c}$ (blue dashed line) and total interference term, $d_{12}^{b\leftrightarrow c} + d_{21}^{b\leftrightarrow c}$ (black dotted line). (b) $d$ (green solid line), $d_{1+2}^{b\leftrightarrow c}$ (blue dashed line), $d_i^{b\leftrightarrow c}$ (red dotted line).

(c) $d$ (green solid line), $d_{11}^{b\leftrightarrow c}$ (blue dashed line), $d_{22}^{b\leftrightarrow c}$ (red dash-dot line), $d_{12}^{b\leftrightarrow c}+d_{21}^{b\leftrightarrow c}$ (magenta dotted line). (d) Populations of the ground (blue dashed line) and second excited (red solid line) states as a function of time when the initial state is a coherent superposition of the ground and excited states with equally weighted populations. (e) Harmonic spectra corresponding to the ground state alone (blue dashed line), and coherent superposition state (black solid line).

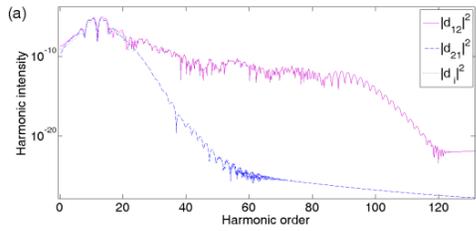

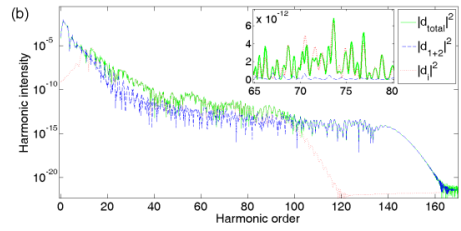

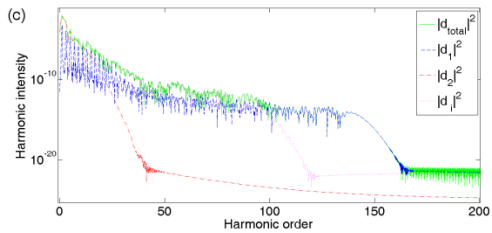

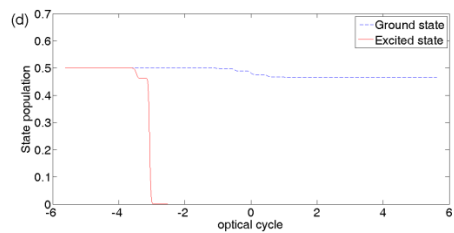

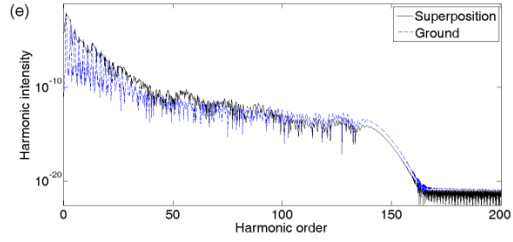

**Figure 4.** Conversion efficiency comparison between the various contributions to the dipole momentum, corresponding to the parameters of figure 2 by using the laser intensity at $I_0 = 1.0e^{15} W/cm^2$. (a) Harmonic intensity as a function of the harmonic order for interference terms $d_{12}^{b\leftrightarrow c}$ (magenta solid line), $d_{21}^{b\leftrightarrow c}$ (blue dashed line) and total interference term, $d_{12}^{b\leftrightarrow c} + d_{21}^{b\leftrightarrow c}$ (black dotted line). (b) $d$ (green solid line), $d_{1+2}^{b\leftrightarrow c}$ (blue dashed line), $d_i^{b\leftrightarrow c}$ (red dotted line); Inset: the interference between the two terms $d_{1+2}^{b\leftrightarrow c}$ and $(d_{12}^{b\leftrightarrow c} + d_{21}^{b\leftrightarrow c})$. (c) $d$ (green solid line), $d_{11}^{b\leftrightarrow c}$ (blue dashed line), $d_{22}^{b\leftrightarrow c}$ (red dash-dot line), $d_{12}^{b\leftrightarrow c} + d_{21}^{b\leftrightarrow c}$ (magenta dotted line). (d) Populations of the ground (blue dashed line) and second excited (red solid line) states as a function of time when the initial state is a coherent superposition of the ground and excited states with equally weighted populations. (e) Harmonic spectra corresponding to the ground state alone (blue dashed line), and coherent superposition state (black solid line).